# An Innovative Approach to Determine Rebar Depth and Size by Comparing GPR Data with a Theoretical Database

Zhongming XIANG, Ph.D. Candidate[1], Ge (Gaby) OU, Assistant Professor[2] and Abbas RASHIDI, Assistant Professor[3]

[1] Department of Civil and Environmental Engineering, University of Utah, Salt Lake City, UT 84112; email: zhongming.xiang@utah.edu

[2] Department of Civil and Environmental Engineering, University of Utah, Salt Lake City, UT 84112; e-mail: ge.ou@utah.edu

[3] Department of Civil and Environmental Engineering, University of Utah, Salt Lake City, UT 84112; e-mail: abbas.rashidi@utah.edu

**ABSTRACT**

Ground penetrating radar (GPR) is an efficient technique used for rapidly recognizing embedded rebar in concrete structures. However, due to the difficulty in extracting signals from GPR data and the intrinsic coupling between the rebar depth and size showing in the data, simultaneously determining rebar depth and size is challenging. This paper proposes an innovative algorithm to address this issue. First, the hyperbola signal from the GPR data is identified by direct wave removal, signal reconstruction and separation. Subsequently, a database is developed from a series of theoretical hyperbolas and then compared with the extracted hyperbola outlines. Finally, the rebar depth and size are determined by searching for the closest counterpart in the database. The obtained results are very promising and indicate that: (1) implementing the method presented in this paper can completely remove the direct wave noise from the GPR data, and can successfully extract the outlines from the interlaced hyperbolas; and (2) the proposed method can simultaneously determine the rebar depth and size with the accuracy of 100% and 95.11%, respectively.

**INTRODUCTION**

Inspecting the status of reinforcing steel bars (rebar) in concrete elements is an important task in structural health monitoring (SHM). Typically, concerns of rebar depth and size matching the design requirement are critical in the concrete structural health. In the last few decades, the automatic methods have been widely used in construction (Sherafat et al. 2019, Taghaddos et al. 2016). For detecting the depth and size of a rebar, several automatic NDT technologies (e.g., covermeter testing, ultrasonic pulse velocity testing, impact echo testing) have been developed and applied. Ground penetrating radar (GPR), one of the NDT methods, is frequently employed to detect the rebar because of its easy handling and rapid scanning.

As the basis of GPR detection, an electromagnetic (EM) wave is generated by GPR and sent into concrete. When reaching the surface of rebar embedded in concrete, the EM wave is reflected and detected by GPR. Through recording the two-way travel time of EM wave, GPR can get the spatial characteristics of the rebar. Normally, the hyperbola pattern is the presentation form of rebar reflection in GPR data. It is combination between the rebar size, depth, the velocity of the EM wave in concrete, scanning direction, and the two-way travel time of the EM wave. Locating the rebar in GPR is already a mature technique (Agred et al. 2018, Xiang et al. 2019a and Wiwatrojanagul et al. 2017). However, it is still a challenge to determine the cover depth and size of rebar in high accuracy with the state-of-the-art approaches (Rathod et al. 2019 and Xiang et al. 2019b).

One practical challenge of GPR is the hyperbola extraction, which concerns the removal of direct wave and identification of hyperbola outlines. On one hand, due to the existence of direct waves in GPR data, parts of rebar signals are covered by the direct wave, and this phenomenon results in the uncompleted hyperbolas. The existing research of direct wave removal can be classified into two directions. The first direction is the subtraction method, which subtracts direct waves from raw data to obtain pure targets signals (Groenenboom et al. 2000, Chantasen et al. 2018 and Mayordomo et al. 2008). The second direction is the wavelet transform method, which focuses on the difference of target signals and direct waves in different domains (Wang et al. 2017 and Shi et al. 2011). However, all of these methods have specific application conditions and complex procedures, thus these methods are difficult to generalize to broad cases. One the other hand, due to the weak signal and the strong noise in GPR data, it is hard to identify hyperbola outlines. The researchers have proposed several different methods to this issue (Lei et al. 2019 and Dou et al. 2016), but all of them extracted the central parts of hyperbolas, which are not the initial rebar reflection in GPR data.

The other practical challenge of GPR is the coupling between the rebar depth and size. As is stated above, the hyperbola of rebar reflection relates to both the depth and size. If adding a minimal change to one variable, the solution of the other will be changed as well. The first attempt to address the coupling issue is assuming the known depth or size of the rebar. In order to calculate the rebar size, Hasan and Yazdani (2016) assumed the cover depth was known, and then compared the maximum amplitude of rebar reflection with different rebar sizes. Conversely, other researchers assume the known rebar size to calculate the depth. For instance, Wiwatrojanagul et al. (2017) assumed the rebar size was zero when estimating depth. They also verified the accuracy was consistent for different rebar sizes. However, these assumptions are not realistic, which limited their applications on the built environment. The second attempt to solve the coupling issue is the modification of GPR devices. For determining the rebar size, Leucci (2012) analyzed the differences of GPR data between co-polarized antennas and cross-polarized antennas. Agred et al. (2018) assembled two receiver antennas and one transmitter antenna. In these cases, one more reference coordinate was developed. Another type of device modification is to combine GPR with other technologies. For example, Zhou et al. (2018) developed a new rebar detection system based on GPR and electromagnetic induction (EMI). Similar to Agred et al. (2018), EMI provided one more reference coordinate compared with a traditional GPR device. However, hardware domain experts are required for these modifications, and thus, the method is cost-prohibitive for real-world application.

The existing research in solving the practical challenges of GPR suffers serious shortcomings. Thus, there is an increasing demand for approaches that can efficiently identify hyperbola outlines and accurately estimate the rebar depth and size at the same time. In order to identify hyperbola outlines, this article proposes an innovative method, which contains an algorithm of direct wave removal and an algorithm of hyperbola reconstruction and separation. Meanwhile, for determining both the depth and size of rebar, a database consisting theoretical hyperbolas is subsequently developed and compared with extracted outlines.

**RESEARCH OBJECTIVES**

The research objectives of this paper are precisely identifying the hyperbola and simultaneously determining rebar depth and size in GPR data, by solving the coupling issue of rebar depth and size in hyperbola. Figure 1 shows four theoretical hyperbolas with sizes of #4 and #5 and depths of 6cm and 7cm. The hyperbolic patterns are affected by rebar size and depth concurrently, creating significant difficulty in simultaneously determining both from the hyperbola patterns. In addition, the initial reflection of rebar signals is the hyperbola outlines, which are the weakest parts in the hyperbolas. Thus, identifying the weak outlines is also a significant issue for determining rebar depth and size.

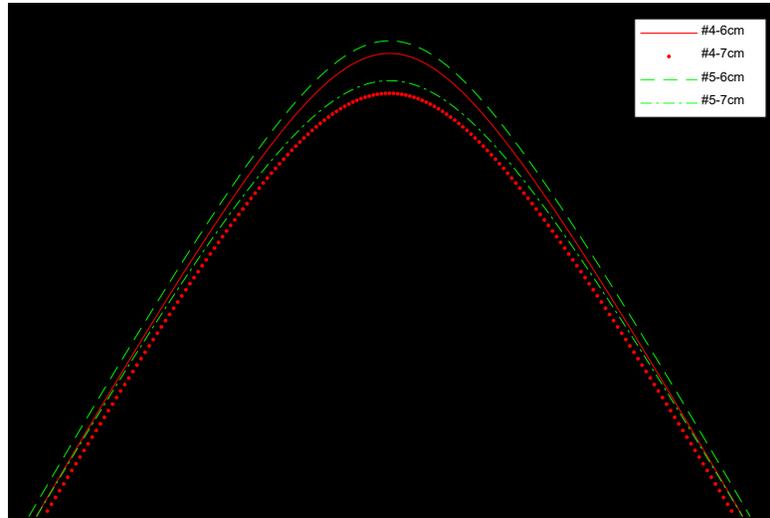

**Figure 1. Theoretical hyperbolas with different rebar sizes and depths**

To address these challenges, this paper proposes an innovative method, the workflow of which is depicted in Figure 2. As the basis of this method, the hyperbolas in GPR data should be identified first. In this step, an algorithm is developed to remove the direct wave and extract the hyperbola from GPR data. And then, considering the discrete rebar sizes and depths in the real construction, several theoretical hyperbolas are generated to form a database. The depth and the size of the rebar target is determined to be the same as its nearest counterpart in the database in terms of the minimum mean distance. More details will be presented in the following chapters.

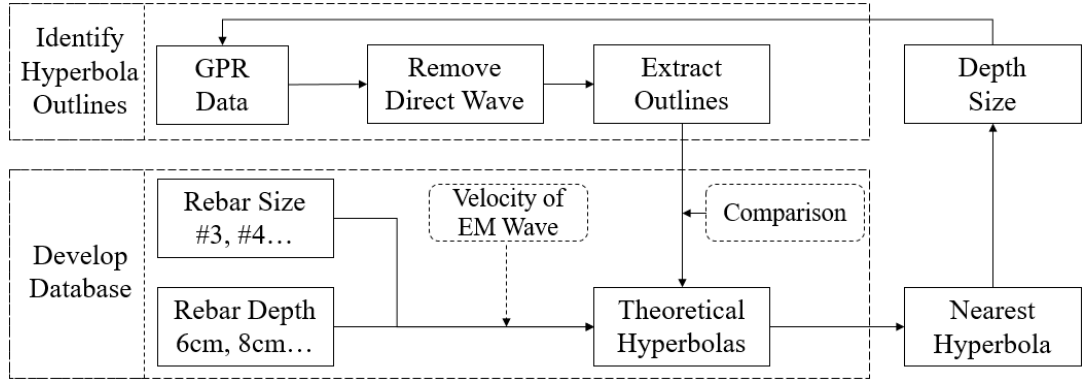

Figure 2. Overall workflow of proposed method

## PROPOSED METHODOLOGY

As noted above, this paper develops a method to compare the hyperbola outlines with the database for simultaneously determining rebar depth and size. For the convenience of explaining the methodology details, a GPR data shown in Figure 3b is synthesized based on a concrete specimen, with its drawing demonstrated in Figure 3a. All of the five rebars have the same diameter of 22.22 mm, but the cover depths from R1 to R5 vary from 6cm to 14cm, with 2cm intervals.

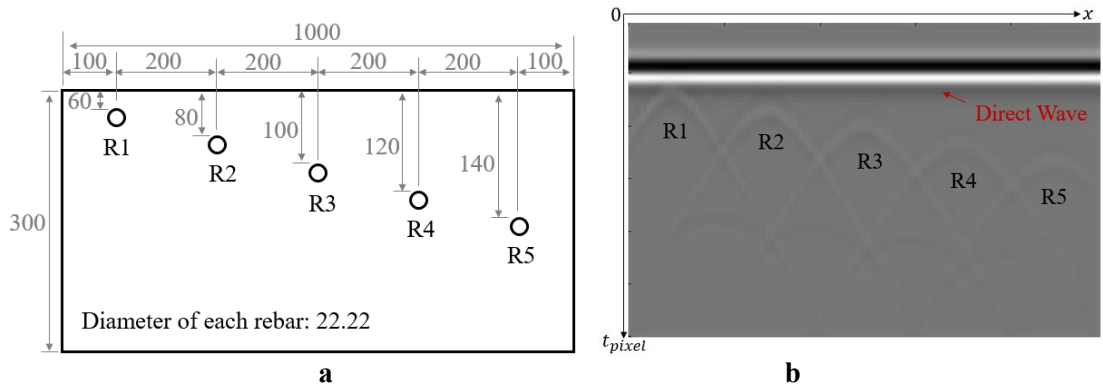

Figure 3. Rebar signal overlapped with direct wave in GPR data: (a) concrete model of rebar arrangement; (b) raw GPR data

### Direct Wave Removal

On real construction facilities, if the rebar depth is shallow, the rebar signal will be overlapped with the direct wave (Figure 3a) in GPR data. As an example, a part of the hyperbola of R1 in Figure 1b is overlapped with the direct wave. In order to extract the hidden rebar signal from the direct wave, an adaptive algorithm is developed based on the subtraction method. There are two steps in this algorithm: GPR data preprocessing and Direct wave subtraction.

*GPR Data Preprocessing*

Since the rebar signal in GPR data is relatively weak, we apply the histogram equalization to enhance the raw GPR data. By adjusting the pixel intensities in the image, histogram equalization increases the contrast of GPR data. After the enhancement, all pixel intensities are mapped to the $t_{pixel}$ axis (pixel domain). By comparing the envelope curve of the mapped intensities with enhanced GPR data, two findings are demonstrated: (1) the pixel intensities of the direct wave are the highest; (2) there is a region with low pixel intensity which is the transition between the direct wave to the rebar reflection. According to the two findings, the raw GPR data can be divided into a rebar reflection region and a direct wave region, and the area to be processed can be concentrated.

*Direct Wave Subtraction*

The details of this step are discussed as follow: (1) the algorithm obtains the most frequent intensity $I_{maxfq}$ in the line, which is the transition from the direct wave region to the rebar reflection region.; (2) a vector $V_p$ is generated to store the pixel positions where the intensities in the same height are $I_{maxfq}$; (3) as shown in Figure 4b, a synthetic direct wave is developed by extending the mean values of intensities $I_{i=1:t_{tp},\ j=V_p}$; (4) and then, the rebar signal that is hidden in the direct wave is extracted by subtracting the synthetic direct wave from raw data (Figure 4a); (5) subsequently, rebar signal is enhanced again with histogram equalization (Figure 4c); (6) finally, the open operation, a kind of noise removal algorithm (Rafael et al. 2018), is applied to remove noise (Figure 4d).

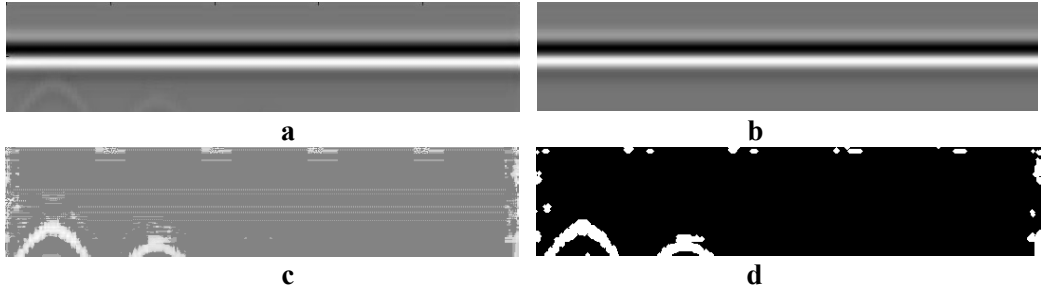

**Figure 4. Direct wave removal process in GPR data: (a) raw GPR data with direct wave; (b) develop the synthetic direct wave; (c) extract the rebar signal; and (d) enhance the rebar signal**

**Hyperbola Outlines Identification**

Through analyzing the graphical features of the rebar reflection region, we have found that the hyperbolas intensities are always higher than the background. Based on this finding, a criterion is developed to remove the background signals:

$$I_{ij}^r > \frac{I_{max}^r - I_{mean}^r}{b} + I_{mean}^r \qquad (1)$$

where $I_{ij}^r$ is the intensity of each pixel in the rebar reflection region, $I_{max}^r$ and $I_{mean}^r$ are the maximum and mean pixel intensity in the rebar reflection region, respectively, and $b$ is a threshold, we use 1.5 in this study. The remaining pixels are shown as white points in Figure 5a. If combining it with the result obtained from the previous step, the hyperbolas without neither direct wave nor background signals are generated (Figure 5b).

In order to extract the full information of one hyperbola, we develop reconstruction and separation steps to further fill the gaps between the discontinuous segments and separate the target hyperbola from other interlaced segments.

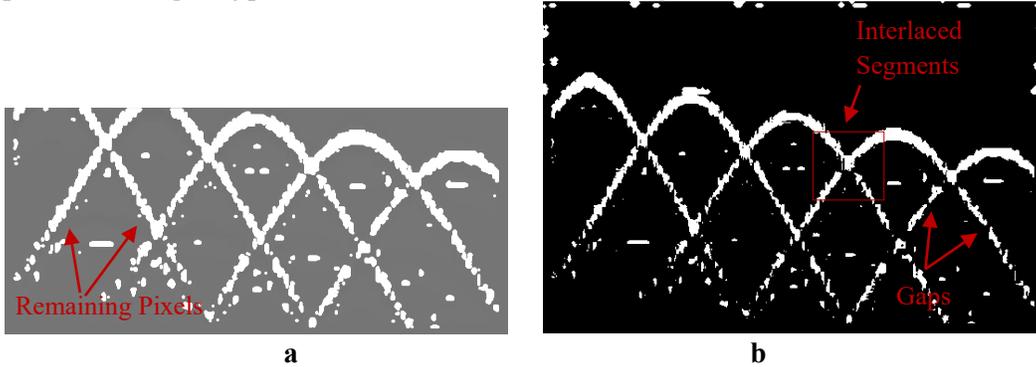

Figure 5. Hyperbolas without direct wave and background: (a) partial view; (b) entire hyperbolas

*Hyperbolas Reconstruction*

To address the issue of discontinuity, this research reconstructs the hyperbolas based on a modified RANSAC algorithm. This algorithm uses the slops of the two paths cross the intersection are almost opposite as the intersection feature (Figure 6a). After that, the intersection points and the fitted lines through these points are generated. If using the fitted lines as bridges, the discontinuous segments near intersections can be connected to form one part. However, there are still some discontinuous segments that are far away from intersections. In order to connect these segments, this research develops a convolutional operation (Rafael et al. 2018), which has two kinds of kernel matrixes corresponding to segments in different directions. The intensity of the central point equals to the product of the mean values of the opposite angles. In this case, only the gaps within hyperbolic patterns are filled (Figure 6a). Similar to the previous steps, noise and unconnected segments are removed by open operation. Figure 6b shows the desired hyperbolas, which are connected as much as possible.

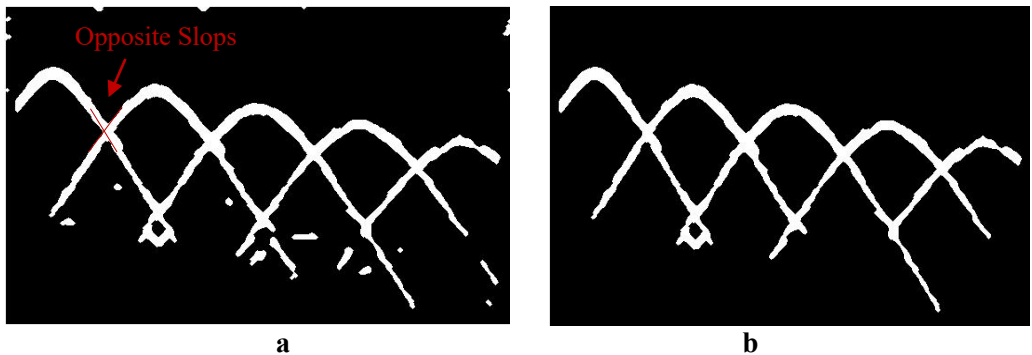

Figure 6. Hyperbolas pattern filled gaps: (a) with noise; (b) without noise

*Hyperbolas Separation*

As shown in Figure 7a, matrixes with the size of $30 \times 30$ are generated at the intersections and applied to erode the hyperbolas. The erosion forms several new segments, which are all labeled in order (Figure 7a) based on the connected domain. And then, through considering the minimum distance between two segments as well as their spatial relations (left, right, up and down), the segments are relabeled to the same number if they belong to one hyperbola. Subsequently, two lines are drawn at the eroded intersections to connect the hyperbolas (Figure 7b). Finally, the hyperbola outlines are extracted and can be used to calculate the rebar depth and size.

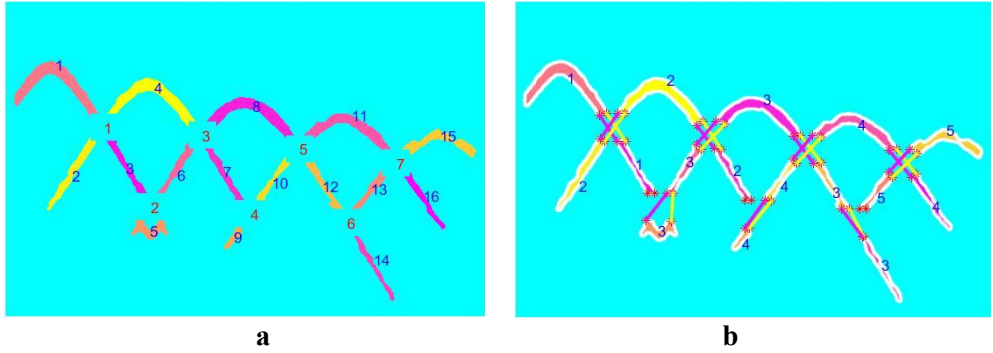

**Figure 7. Hyperbola separation: (a) labeled segments; (b) reassembled segments**

**Database Construction**

To address the aforementioned coupling issue, this research proposes a hyperbola comparison approach to a theoretical database. According to James (2016), the types of cover depths and rebar sizes are consistent, therefore, there is no need to consider any unusual case when determining the rebar depth and size on real buildings. If assuming that velocity of EM wave is a constant, the discrete depths and sizes can be combined to simulate several theoretical cases. The hyperbola of each cases can be developed, which are the elements of the database. For finding the best element, the algorithm calculates the mean value of the distances from all pixels in the experimental hyperbola to the element in the database at first. And then, among the database, the element corresponding to the minimum mean value of distances is considered as the nearest one. After the nearest element is determined, the corresponding depth and size are assigned to the experimental hyperbola. The workflow of determining rebar depth and size is shown in Figure 8.

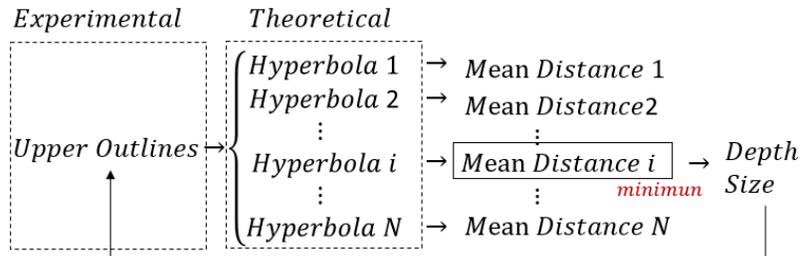

**Figure 8. Workflow of determining rebar depth and size**

**EXPERIMENT AND RESULTS**

Figure 9 shows the experimental setup of a concrete specimen with five rebars. In order to make the experiment similar to real construction case as much as possible, this research selects regular rebar sizes, which are #3, #4, #5, #6, #7, #8, #9, #10, #11,

#14 and #18. The nominal diameters of these rebars can be referred from James (2016). Meanwhile, the depths are selected as 6cm, 8cm, 10cm, 12cm and 14cm. Each rebar size is tested with each depth except some cases that rebar sizes are too large compared with the depth. There are five rebars in each case, as shown in Figure 10a. Table 1 depicts the details of the 45 cases, and there are 225 rebars considered in the entire study. The GPR data are synthesized by gprMax, an electromagnetic simulation software. The central frequency of which is 1.5 GHz. Meanwhile, the relative permittivity of the concrete is 6.0.

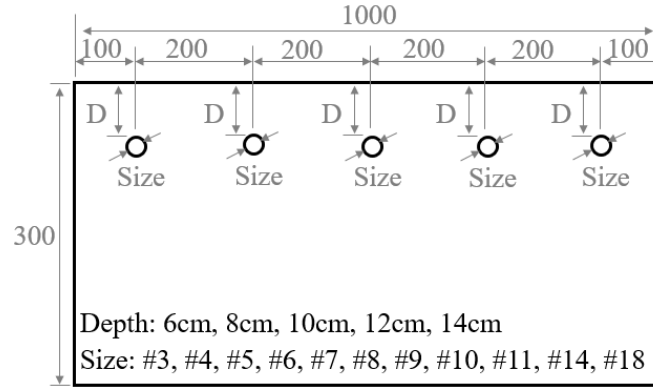

Figure 9. Concrete specimen with different rebar depths and sizes

Table 1. Size and depths of rebars considered in each case

| Case No. | 1-6 | 7-15 | 16-24 | 25-34 | 35-45 |
|---|---|---|---|---|---|
| Depth | 6cm | 8cm | 10cm | 12cm | 14cm |
| Size | #3-#8 | #3-#11 | #3-#11 | #3-#14[1] | #3-#18[2] |

Note: 1. There is no #12. 2. There is no #12, #13, #15, #16 and #17.

Applying the proposed method to each cases, the results of estimating rebar depths and sizes are obtained and shown in Table 2. For rebar depth determination, 225 out of 225 rebars are correctly estimated, thus, the accuracy is 100%. It means the proposed method is good at rebar depth estimation. For rebar size determination, 214 rebars are correctly determined as their true sizes, and 11 rebars are determined incorrectly. The accuracy of rebar size estimation reaches to 95.11%. Moreover, if analyzing the incorrectly estimated cases, it's found that all of these rebar sizes are predicted to the neighbor size. In other words, the proposed method can determine the rebar size at high accuracy, and even the incorrect estimations are in a rational range.

Table 2. Estimated results of rebar depths and sizes

| Items | Total Number | Correctly Estimation | Incorrectly Estimation | Accuracy | Accuracy +/- 1 interval |
|---|---|---|---|---|---|
| Depth | 225 | 225 | 0 | 100% | 100% |
| Size | 225 | 214 | 11 | 95.11% | 100% |

**CONCLUSION**

The accurate estimation of rebar size and depth in concrete using GPR data is a significant challenge faced by many engineers, especially those involved in structural health monitoring. On one hand, the existence of direct waves, the weak reflection of rebar signal, and the interlaced hyperbolas increase the difficulty of extracting hyperbolas from GPR data. Furthermore, even with a clear hyperbola, the coupling between the rebar depth and size makes it difficult to simultaneously determine these two variables. This paper has examined an innovative method that compares experimental data with the database of theoretical hyperbolas in order to simultaneously determine rebar size and depth from GPR data. As the basis of this method, an adaptive algorithm has been developed to remove direct waves. Meanwhile, this paper identifies the hyperbola outlines, which can be directly used to estimate the rebar properties. Subsequently, several theoretical hyperbolas were simulated to construct a database. The distances of the target hyperbola to the outlines are compared to determine its closet representation in the database. The depth and size of the most similar value is assigned to experimental rebar. Finally, 45 concrete elements with 225 rebars are simulated by gprMax and tested with the proposed method. The main understandings from the results are as follows:

- The coupling of rebar depth and size is successfully solved by the proposed method, which can simultaneously determine rebar depth and size by comparing GPR data with the database of theoretical hyperbolas.
- The rebar signals covered by direct waves are completely extracted. Moreover, this research can precisely identify the hyperbola outlines from the weak rebar signals in GPR data.
- The proposed method shows rebar depth estimation proficiency, with an accuracy of which can reach 100%. It is also very robust based on validation in cases of different rebar sizes and depths.
- The proposed method can determine rebar size at an accuracy of 95.11%. Even for the incorrect cases, the estimation results are incredibly close in size.

However, as a limitation, this paper assumes that the velocity of the EM wave is a known parameter. In the future, the authors plan to estimate the velocity of the EM wave as well. In addition, the validation is implemented on simulation data. Thus, the authors plan to repeat this paper on real data.